\documentclass{article}

\PassOptionsToPackage{numbers, compress}{natbib}

  \usepackage[preprint]{neurips_2019}
\usepackage{times}

\usepackage[utf8]{inputenc} 
\usepackage[T1]{fontenc}    
\usepackage{hyperref}       
\usepackage{url}            
\usepackage{booktabs}       
\usepackage{amsfonts}       
\usepackage{nicefrac}       
\usepackage{microtype}      
\usepackage{amsmath}
\usepackage{amssymb}
\usepackage{graphicx}
\usepackage{multirow}
\usepackage{color}
\usepackage{arydshln}

\title{EncryptGAN: Image Steganography with Domain Transform}

\author{%
  Ziqiang Zheng\\
  Ocean University of China\\
  Qingdao, China\\
  \And
  Hongzhi Liu \\
  Ocean University of China\\
  Qingdao, China\\
  \AND
  Zhibin Yu \\
  Ocean University of China\\
  Qingdao, China\\
  \And
  Haiyong Zheng \\
  Ocean University of China\\
  Qingdao, China\\
  \And
  Yang Wu \\
  Nara Institute of Science and Technology\\
  Nara, Japan\\
  \And
  Yang Yang \\
  University of Electronic Science and Technology of China\\
  Chengdu, China\\
  \And
  Jianbo Shi \\
  University of Pennsylvania\\
  Philadelphia, US\\
}

\begin{document}

\maketitle

\begin{abstract}
We propose an image steganographic algorithm called EncryptGAN, which disguises private image communication in an open communication channel. The insight is that content transform between two very different domains (e.g., face to flower) allows one to hide image messages in one domain (face) and communicate using its counterpart in another domain (flower). The key ingredient in our method, unlike related approaches, is a specially trained network to extract transformed images from both domains and use them as the public and private keys. We ensure the image communication remain secret except for the intended recipient even when the content transformation networks are exposed.  

To communicate, one directly pastes the `message' image onto a larger public key image (face). Depending on the location and content of the message image, the `disguise' image (flower) alters its appearance and shape while maintaining its overall objectiveness (flower). The recipient decodes the alternated image to uncover the original image message using its message image key. We implement the entire procedure as a constrained Cycle-GAN, where the public and the private key generating network is used as an additional constraint to the cycle consistency. Comprehensive experimental results show our EncryptGAN outperforms the state-of-arts in terms of both encryption and security measures. Our code will be available at \url{https://github.com/zhengziqiang/EncryptGAN}
\end{abstract}

\section{Introduction}

Image steganographic techniques conceal private image message within another image. The goal is to make the concealed images appear to be something else. One approach for concealment is akin to writing with invisible ink on a `cover' image.  These steganographic techniques modify `cover' images for hiding information including spatial domain~\citep{li2011survey,bender1996techniques,cheddad2010digital,juneja2009designing,bailey2006evaluation,naghsh2006new}, transform domain~\citep{johnson2000survey,wang2004cyber,kharrazi2007image,hamid2012image}, spread spectrum~\citep{anderson1998limits,smith1996modulation,alturki2001secure,youail2007improved,singh2010spread}, statistic methods~\citep{weiss2009principles,radhakrishnan2002data}, distortion techniques~\citep{reddy2009high}, image generation~\citep{kruus2003survey}, element modification and adaptive steganography~\citep{fridrich1999applications}. For many of the cases, current methods either leave a visible trace of the hidden image, or it is sensitive to corruption of the `covered' image.

A new idea of image steganography was proposed based on domain transform~\cite{chu2017cyclegan}.  The main idea is the following. Instead of focusing on invisible inks, we will write with visible ink (private image) on visible text of one language (Cover image), and translate the combined text to another language (Disguise image).  The translation process is highly nonlinear with respect to how the two written texts are interacting. A carefully designed reverse mapping function is used to reconstruct the original message and separate out the cover image with private image. 


The main weakness of this approach is that it is a symmetric encryption mechanism. The two parties of the communication depends the security of the translation mapping functions: $\mathcal{F}$ for encryption and $\mathcal{G}$ for decryption. Once the encryption function $\mathcal{F}$ and decryption function $\mathcal{G}$ is known, the communication becomes public. In addition, a symmetric image concealment method could not perform digital signature: the two communicating parties do not know if they are communicating with the person they wish to communicate with. We build on this insight of using domain transformation as image concealment and provides strong privacy protection using asymmetric encryption mechanism, that includes a pair of public key and private key.   

The key innovative ingredient of our method is a specially learned network that transforms images from the two domains into secure public and private keys that are integrated into the encryption and decryption learning process.

\begin{figure}[!t]
	\centering   
	\includegraphics[width=\linewidth]{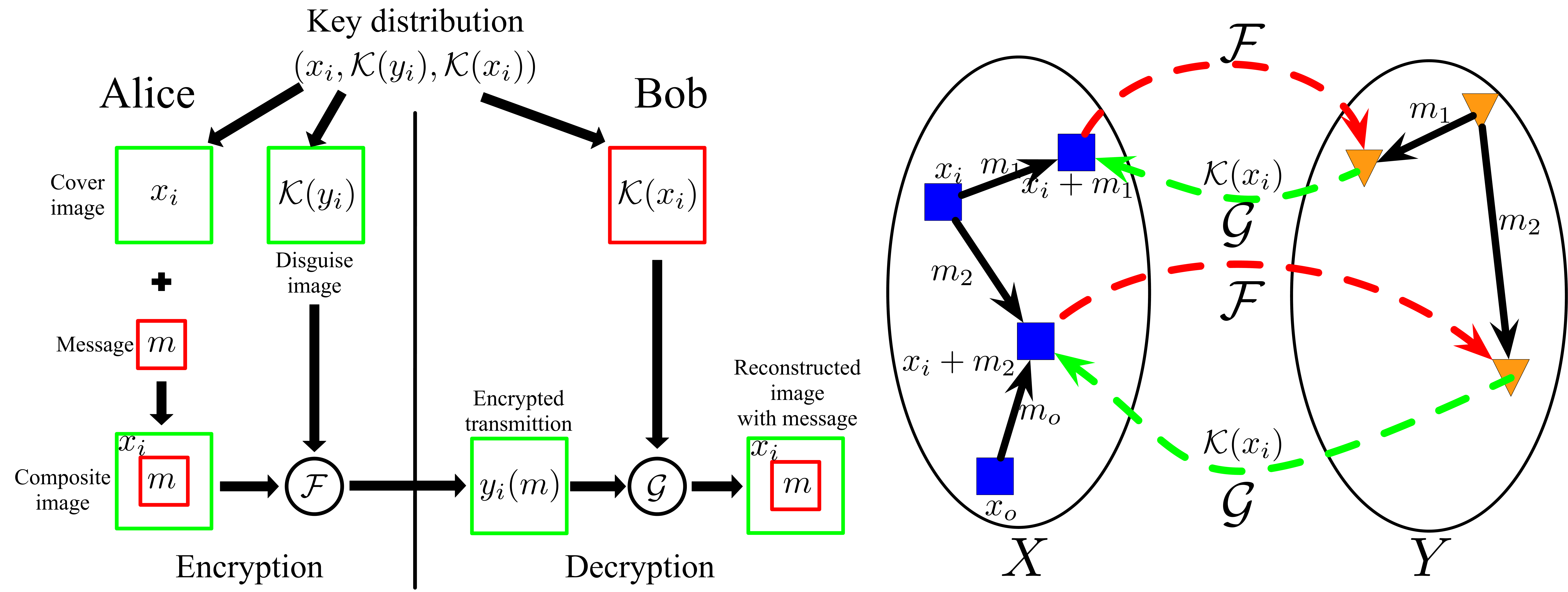}
	\caption{{\bf Left:} EncryptGAN asymmetric encryption image steganographic algorithm flowchart.  Boxes in \textcolor{green}{green} denote public information, boxes in \textcolor{red}{red} contain private key or messages.  Two distinct image domains provide two random images including a `cover' image $x_i$ and a `disguise' image $y_i$.  Before the communication, a third party transformed them via a learned network $\mathcal{K}$ into unrecognizable images: sender is given $\mathcal{K}(y_i)$ (public) and receiver is given $\mathcal{K}(x_i)$ (private). Sender pastes in the private message $m$ on the `cover' and using the transformed `disguise' $\mathcal{K}(y_i)$ to create an altered version of `disguise' image $y_i(m)$.  Receiver decodes the altered `disguise' image with the transformed `cover' image $\mathcal{K}(x_i)$ (private) back to the original private message.
	{\bf Right:} An intuitive explanation of our steganographic operation principle as navigation on manifolds of two image domains.   A `cover' image $x_i$ when pasted with different private messages, $m_1, m_2$, induces movements to distinct landmarks on its image manifold $X$. The Cycle-GAN generative networks ensure similar movements on manifold $Y$ relative to the `disguise' image landmark. To decode the original private message, it is crucial to know the original reference landmark of the `cover' image $x_i$, because an alternative `cover' image $x_{o}$ can also reach the same landmark location: $x_{o} + m_{o}$ = $x_{i} + m_{2}$. This multiplicity of the solution provides strong privacy protection.  We conjunct that $\mathcal{K}(x_i)$ needs to encode just enough information to infer the landmark location of $x_{i}$ to compute $m_{2}$. } 
	\label{fig:illustration} 
\end{figure}

We call our asymmetric image encryption algorithm EncryptGAN, which combines adversarial training in a constrained Cycle-GAN framework. 
Imagine a situation as illustrated in Figure~\ref{fig:illustration}, Alice wants to send a private image (for example her cat) to Bob.  First, a trusted third party pick a pair `cover' and `disguise' images from two distinct image domains.  For example, the `cover' image could be a celebrity face, and `disguise' a flower.  The two images are transformed into a pair of public and private image keys using a specially trained network.
Alice is given a `cover` and transformed `disguise' image as Bob's public key, Bob is given a transformed `cover' image as his private key.  Alice pastes in her private image (cat) into the `cover' image (face) and uses Bob's public key to generate an altered flower image that resembles the `disguise' image.  Bob decrypts the received image using his own private key image to recover the image of `face + cat'. Even when the complete functions of encryption and decryption are leaked, we can still ensure the secrecy of the transmitted image.

Our EncryptGAN can also achieve digital signature. If Alice needs to make sure that she is communicating with Bob, Bob can use the secret image as his own private key to obtain an encrypted image and send the encrypted image to Alice.  Alice decodes the encrypted image with Bob's public key and implements the secret matching using her own secret image. If the images do not match, Alice knows this information is not from Bob, hence achieving the digital signature functionality. 

\section{Proposed Approach}

\subsection{Framework}
\begin{figure}
	\centering   
	\includegraphics[width=\textwidth]{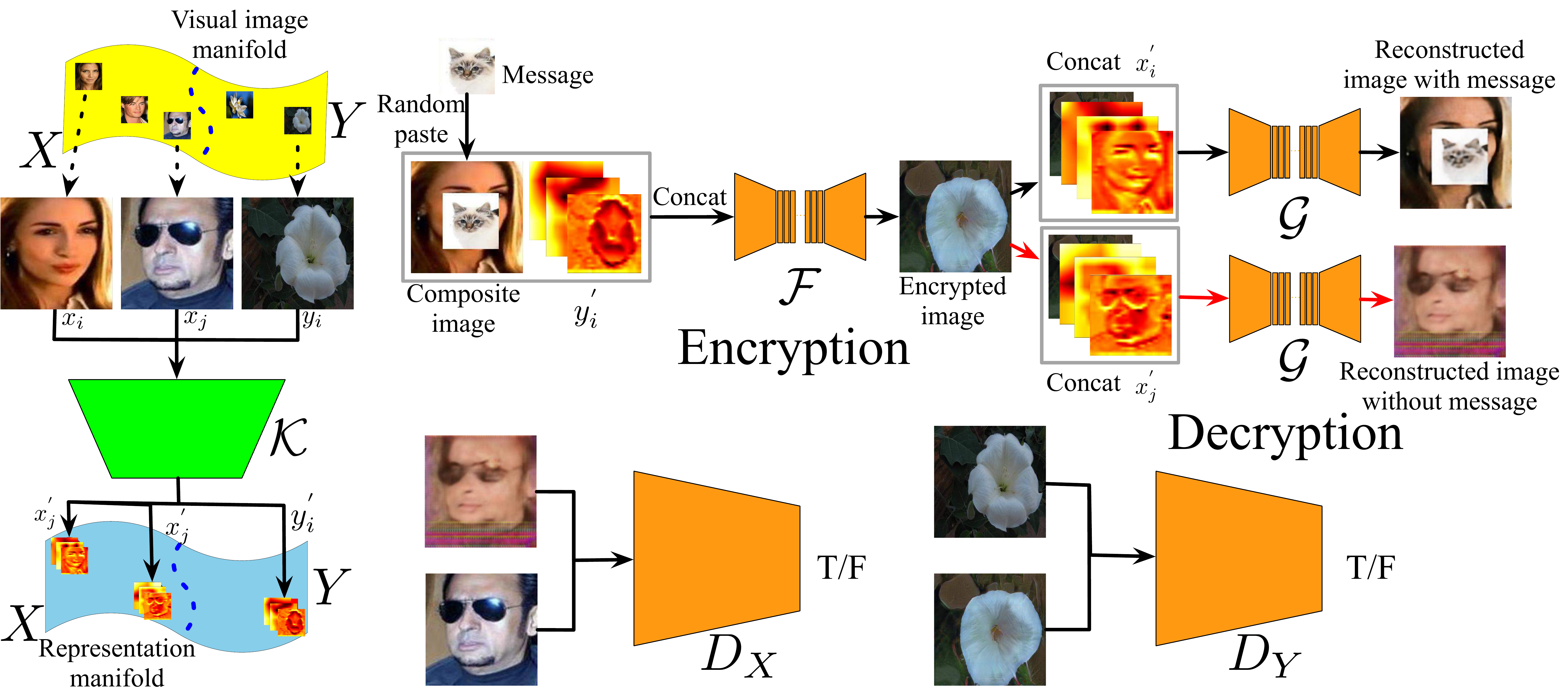}
	\caption{The network architecture of our EncryptGAN. }
	\label{fig:archi} 
\end{figure}

Figure~\ref{fig:archi} shows the overall design of our EncryptGAN. We build on the architecture of CycleGAN~\cite{zhu2017unpaired}, including two generators: $\mathcal{F}$ and $\mathcal{G}$, and two discriminators ${D_{x}}$ and ${D_y}$, which are responsible for translating images between two domains $X$ and $Y$. 

To provide strong privacy protection, asymmetric encryption mechanism~\cite{bellare1994optimal}, requires a public key and a private key for encryption and decryption. 

Our main innovation is creating a specially designed network, a key generator $\mathcal{K}$, for transforming images from both domains into unrecognizable public-private key-images.  In our experiment, we used a 6-layers convolutional neural network for $\mathcal{K}$ and constrained the output to be a 3-channel image.

A trusted third party randomly choose two images $x_i$ and $y_j$ from two image domains $X$ and $Y$, and using $\mathcal{K}$ to build a private key $x_i'$ from $x_i$, and a public key $y_j'$ from $y_j$.

To transmit a message, we use the $x_i$ as the `cover' image and paste in our private message image $m$ at a random position of $x_i$ to create a composite image. During the encryption process, the `cover' image is concatenated with the public key $y_j'$ to feed the encoder $\mathcal{F}$ for generating an encrypted image $y_i(m)$. In the decryption stage, both the private key $x_i'$ and the encrypted image $y_i(m)$ are used as the input to the decoder $\mathcal{G}$ for decryption. 


A key property we would like to demonstrate is that only the correct private key image could lead to the correct reconstruction of the `cover' image plus the message image.

\subsection{Key Pairs Generation using Disentangled Representation}
 A deep neural network typically extracts semantic representations in higher layers and low-level detailed information in shallow layers~\cite{brahma2015deep}. By carefully choosing which level features to keep and remove in different layers, we create an irreversible process, which provides candidates for key generation. Following this idea, we adopt an independent convolutional neural network $\mathcal{K}$ as a key generator for our EncryptGAN to create the key pairs (public and private keys).

There are many reasons to prefer asymmetric encryption to symmetric encryption~\cite{fujisaki1999secure}. Asymmetric encryption mechanism based a public key and a private key provides an extra layer of privacy protection for the sender and the receiver. To generate diverse key pairs for asymmetric encryption, we define a key matching loss to encourage the encoded image to have similar feature representations with the source of the keys as follows:
\begin{equation}
\begin{split}
\mathcal{L}_{key}(\mathcal{F},\mathcal{G},x) =\sum_{i=1}^{N} \mathbb{E}_{x\thicksim P_{data(x)}}[||L_{i}(\mathcal{G}( \mathcal{F}(x)))-L_{i}(x)||_{1}+||L_{i}(\mathcal{F}( \mathcal{G}(x)))-L_{i}(x)||_{1}]
\end{split}
\label{q1}
\end{equation} 
where $L_{i}(x)$ is the activation response of the $i_{th}$ layer from $\mathcal{K}$ with the input $x$; $\mathcal{F}(x)$ is the recovered image generated from $\mathcal{G}$; $N$ is the number of layers we need for key generation. We choose $N=3$ layers ($L_{3}$,$L_{5}$ and $L_{6}$) in our experiments. The key matching loss aims to keep the representation similarity between $x$ and $\mathcal{G}(\mathcal{F}(x))$ as well as $y$ and $\mathcal{F}(\mathcal{G}(y))$. 

\subsection{Encryption and Security}
One elemental requirement of an encryption mechanism is to hide information during the encryption stage and recover the information in the decryption stage. 

\textbf{Information Loss.} To ensure the information quality of the recovered image, we introduce the information loss to `cover' the message information. The information loss $\mathcal{L}_{loc}$ is described as:
\begin{equation}
\begin{split}
\mathcal{L}_{info}(\mathcal{F},\mathcal{G})=&\mathbb{E}_{x_{info}\thicksim P_{data(x_{info})}}\parallel \mathcal{G}(\mathcal{F}(x))-x_{info}\parallel_1
+ \mathbb{E}_{y_{info}\thicksim P_{data(\bar{y})}}\parallel \mathcal{F}(\mathcal{G}(y))-y_{info}\parallel_1
\end{split}\label{q2}
\end{equation} 
where $x_{info}$ and $y_{info}$ denote the local area including message information on both composite image and recovered image. The information loss aims to preserve the message information in the decrypted images. 

Security is another basic requirement of an encryption mechanism. When we try to decrypt the encrypted image with an incorrect private key, we will obtain an image without correct message. Furthermore, the output images with wrong keys should not leave any cues of the `cover' images, which would be the correct private key for decryption.

\textbf{Security Loss.} During the reconstruction process, to perform the asymmetric mechanism, we use two different visual images to obtain two different private keys (a true private key and a random private key) to reconstruct the encrypted images.
There are two possible situations during decryption. We hope the decrypted image with a true private key should recover the contents correctly, while the decrypted image with an incorrect private key should fail to obtain the information as well as the `cover' image. During training, we randomly choose a correct private key and an incorrect private key with a probability of 50 percents, respectively. In the case of a correct private key, we use the cycle consistency loss along with the adversarial loss to increase the similarity between the recovered image and the composite image. Otherwise, we optimize the model only using adversarial loss. The security loss is described as:
\begin{equation}
\begin{split}
\mathcal{L}_{s_{correct}}(\mathcal{F},\mathcal{G}) &=\mathcal{L}_{cyc}+\mathcal{L}_{adv}\\
\mathcal{L}_{s_{incorrect}}(\mathcal{F},\mathcal{G}) &= \mathcal{L}_{adv}
\end{split}
\label{q3}
\end{equation}
\begin{equation}
\mathcal{L}_{cyc}(\mathcal{F}, \mathcal{G})= \parallel \mathcal{G}(\mathcal{F}(x))-x\parallel_1+\parallel \mathcal{F}(\mathcal{G}(y))-y\parallel_1
\end{equation}
\begin{equation}
\mathcal{L}_{adv}(\mathcal{F}, \mathcal{G})=\mathbb{E}_{x\thicksim P_{data}(x)}[(\mathcal{D}(x))-1)^2]+\mathbb{E}_{x\thicksim P_{data}(x)}[(\mathcal{D}(G(x)))^2]
\label{q4}
\end{equation} 
where $\mathcal{L}_{s_{correct}}$ and $\mathcal{L}_{s_{incorrect}}$ denote the security loss with a correct/incorrect private key respectively. Please note that the cycle-consistency loss~\cite{zhu2017unpaired} is only adopted between the recovered image and the composite image when a correct private key is given. There is no need to use consistency loss when an incorrect private key is used because the recovered image should not be similar to the input image at this moment. In both cases, the least square adversarial loss~\cite{mao2016multi} is used to restrain the reconstructed image in the same domain as the composite image on semantic level.






\section{Experiments}
\label{experiment}
\subsection{Implement Details}
Our backbone network derives from CycleGAN~\citet{zhu2017unpaired} using 9 residual blocks in the bottleneck of the generators. Our key generation $\mathcal{K}$ has 6 \texttt{Convolution-Relu-Instance} modules, followed with a fully connected layer, and all the kernel sizes are set to 4. $\mathcal{K}$ is optimized by cross-entropy loss to classify which domain the real images are from. We apply the final objective loss to the generator with Adam optimizer of learning rate $0.0002$. Besides, we also impose some additional noise to the input for improving the robustness of model. We conduct experiments on two irrelevant image domains: human face images from \textbf{CelebA}~\cite{liu2015deep} for $X$ and flower images from \textbf{102 Oxford Flower}~\cite{Nilsback08} for $Y$, and all images are resized to 256$\times$256. For message images, we use the images from \textbf{SVHN}~\cite{netzer2011reading} and \textbf{Cat2dog}~\cite{lee2018diverse} datasets. 

We compare our EncryptGAN with available state-of-the-art methods including CycleGAN~\cite{zhu2017unpaired}, MUNIT~\cite{huang2018multimodal}, DRIT~\cite{lee2018diverse}, and Steganography~\cite{baluja2017hiding}. More details and results are shown in \emph{supplementary file} for reference.

\begin{figure}[!t]
	\centering   
	\includegraphics[width=\textwidth]{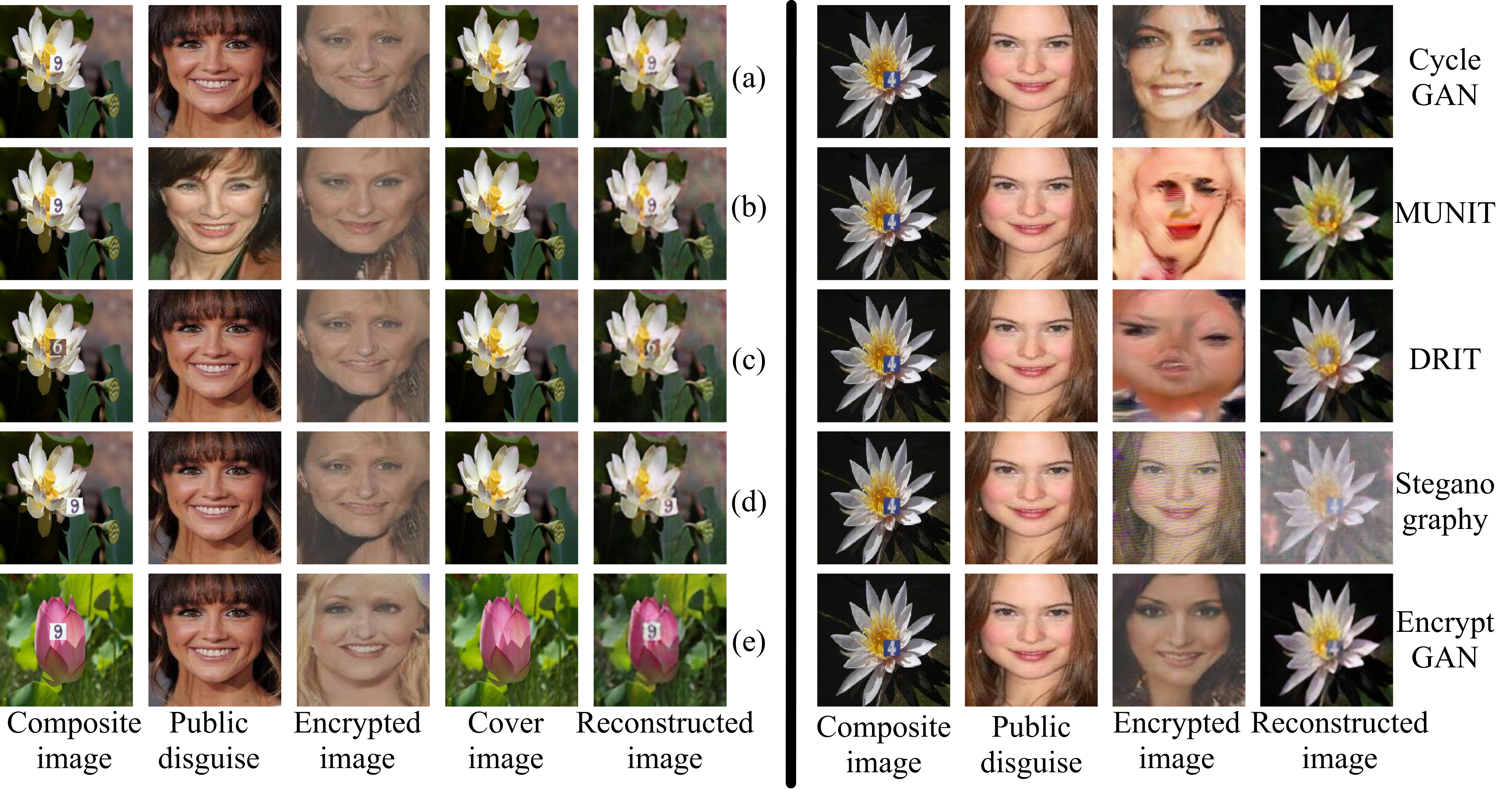}
	\caption{{\bf Left:} The effect under different circumstances. (a) and (b) are different on public disguise. We choose different message digits in (a) and (c). Besides, position changes in both (a) and (d) illustrate the randomness of encrypted information location. As to (e), we achieve satisfying encryption on different cover images.  {\bf Right:} Visual comparison results of different methods.}
	\label{fig:results} 
\end{figure}

\begin{table}
	\caption{Quantitative comparison using different methods. We don't perform security evaluation for other methods because they are not public-key systems.}
	\label{table:comparison}
	\centering
    \resizebox{\textwidth}{!}{
	\begin{tabular}{l c c c c c c c c}
		\toprule
		Methods &  FID $\downarrow$ & Encryption(LPIPS $\uparrow$ / MS $\uparrow$) & MSE $\downarrow$ & RMSE $\downarrow$ & PSNR $\uparrow$ & SSIM $\uparrow$ & LPIPS $\downarrow$ & Security(LPIPS $\uparrow$ / MSE $\uparrow$)  \\
		\midrule
		CycleGAN   & 160.9643  & 0.4510 / 0.1039 & 0.0332 & 0.1769 & 14.3246 & 0.9654 & 0.3115 & - \\
		MUNIT & 218.6034  & 0.3937 / \textbf{0.1308} &  0.0464 & 0.2122 & 12.6177 & 0.9437 & 0.4341 &- \\
		DRIT  & 165.0640 & \textbf{0.4707} / 0.0916 & 0.0492 & 0.2111 & 12.9483 & 0.9206& 0.3174 & - \\
        \hdashline
        Steganography & 238.6737 & 0.3583 / 0.1020 & 0.0225 & 	0.1426 & 16.1933 & 0.9473 & 0.2429& -\\
        \hdashline
		EncryptGAN & \textbf{107.9238} & 0.4141 / 0.0765  & \textbf{0.0139} & \textbf{0.1156} & \textbf{17.5992} & \textbf{0.9825} & \textbf{0.2414} & 0.5191 / 0.1379\\
		\bottomrule
	\end{tabular}
	}
\end{table}

\subsection{Evaluation Metrics}
\noindent\textbf{Encryption performance} is necessary for image steganography. To design a fair comparison, we compute perceptual similarity using the part regions to place message cropped from encrypted and composite images (`cover' + message). For this, we compute Learned Perceptual Image Patch Similarity (LPIPS~\cite{zhang2018perceptual}) and Mean Square Error (MSE) between encrypted and composite images, and denote as \textbf{Encryption(LPIPS/MSE)}. Higher LPIPS and MSE indexes represent better encryption performance and more difficult to decode the message. Besides, Fr\'echet Inception Distance (\textbf{FID}~\cite{heusel2017gans}) computes the distance between the real sample distribution and the encrypted sample distribution. Lower FID score indicates higher image generation quality.

\noindent\textbf{Decryption performance} is also important to ensure how much we can decode the true message. More evaluation metrics are required to evaluate the distance between input message image and reconstructed message image (cropped from the whole image) after decryption, where we compute \textbf{PSNR}, \textbf{MSE}, \textbf{RMSE}, \textbf{SSIM} and \textbf{LPIPS}. Lower MSE and RMSE scores show better reconstruction performance. PSNR can roughly evaluate image quality independently, and usually higher PSNR indicates better performance, while higher SSIM represents higher structural similarity. Note, we compute above 5 metrics between the decoded image and input image.

\noindent\textbf{Security performance} is necessary for encryption, to indicate how much others cannot decode the true message using wrong private keys. Here we compute the average \textbf{difference distance} between message and decoded message with other all wrong private keys generated from testing data. Note, we compute the LPIPS and MSE losses between message image and decoded message images, and denote as \textbf{Security(LPIPS/MSE)}. Higher difference distance represents better security performance.

\subsection{Simple Information Encryption}
In this section, we perform experiments using facial images cropped from CelebA (domain $X$) and flower images from 102 Oxford Flower (domain $Y$). We aim to encrypt message images from SVHN. For this task, we also conduct experiments using CycleGAN~\cite{zhu2017unpaired}, MUNIT~\cite{huang2018multimodal}, DRIT~\cite{lee2018diverse}, and Steganography~\cite{baluja2017hiding} in fair setting. The visual comparison results are exhibited in Figure~\ref{fig:results}. Compared to image deep steganography~\cite{baluja2017hiding}, our method can achieve better image encryption performance visually. Other symmetric image-to-image methods cannot reconstruct the message information, which fail to generate plausible results between two irrelevant domains. Besides, the quantitative comparison among different methods can be found in Table~\ref{table:comparison}. Our EncryptGAN obtains lowest LPIPS and highest SSIM scores, indicating best reconstruction performance. Besides, our method can generate more plausible images with lowest FID.

\begin{figure}[!t]
	\centering   
	\includegraphics[width=\linewidth]{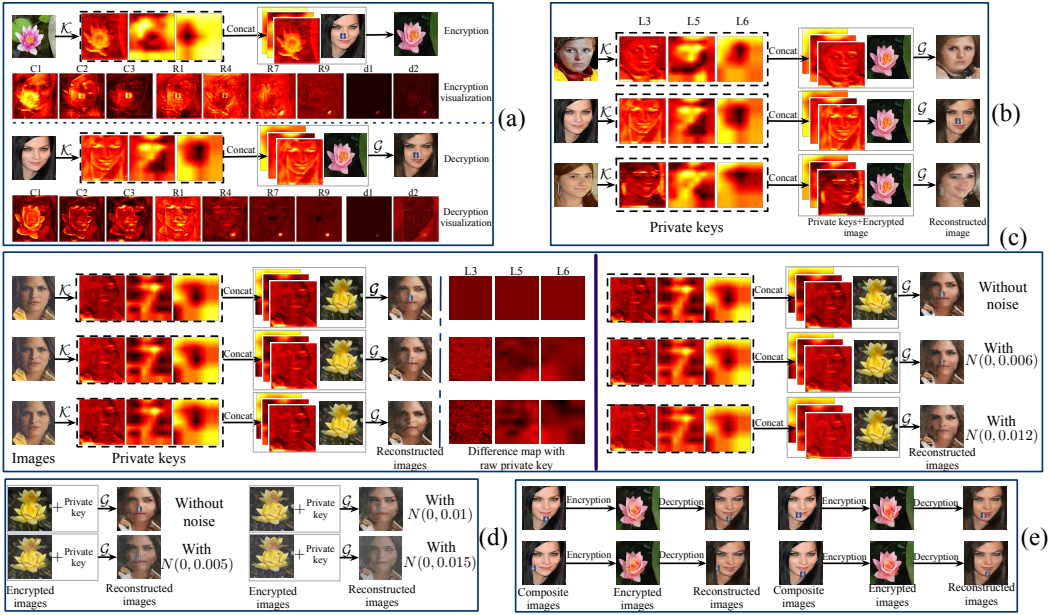}
	\caption{(a) The visual action response results of different layers from generators for both encryption and decryption stages. (b) The effectiveness of the specific private key. (c) The exploration of key sensitivity of our EncryptGAN. (d) The exploration of robustness of our EncryptGAN. (e) Our EncryptGAN can encrypt message at random position.}
	\label{fig:visual} 
\end{figure}

To reveal how the encryption function $\mathcal{F}$ hides the secret information in the encryption stage, and how the decryption function $\mathcal{G}$ decodes the required information, we visualize activation responses of each layer from two generators, and find the information is actually encrypted in the residual blocks. The visualization results are shown in Figure~\ref{fig:visual}(a), we can see that the message information is encrypted at \texttt{R7} block during the encryption. For the decryption stage, it is shown that we obtain the message information at \texttt{R7} block. In order to show the effectiveness of the private key, we use different private keys generated from other random images to recover the encrypted image. As shown in Figure~\ref{fig:visual}(b), only our specific private key can decode the required message information correctly. 

We devise experiments to evaluate the key sensitivity of our EncryptGAN by adding different scales of Gaussian noise. Here we define two cases: 1) we add noise to the visual image and generate the corresponding private key, 2) we directly add different levels of noise to the private key. For the first case, we visualize the difference maps between the raw private key and corresponding private keys generated from images with different scales of noise. We also exhibit the reconstruction performance at different settings. Results are shown on the left and right of Figure~\ref{fig:visual}(c). As shown, if we add noise more than $N(0,0.012)$, our EncryptGAN fails to reconstruct the message information for the first case. The difference maps show that the key generation function $\mathcal{K}$ is sensitive to the noise and $\mathcal{K}$ magnifies the difference between samples, which can help to make the key space sparse. For the second case, we see that our EncryptGAN fails to decode message information if the noise is beyond $N(0,0.012)$. Our model achieves high key sensitivity, and a small perturbance can change the reconstruction output significantly. Besides, we also explore the robustness of our method by adding noise to the encrypted image in Figure~\ref{fig:visual}(d), we fail to decode message from encrypted image with noise beyond $N(0,0.015)$. Finally, we also explore the effectiveness of the position to place the message information. For this task, we randomly place the message images at different regions, and the visual results are shown in Figure~\ref{fig:visual}(e). As shown, our EncryptGAN can still reconstruct the message information no matter where it is, indicating that our model is insensitive to the location of the message images. In other words, EncryptGAN is sensitive to noise while insensitive to location, that is, our method is secret yet robust with a good generalization performance.

\subsubsection{Ablation Study}
\begin{figure}[!t]
	\centering   
	\includegraphics[width=\textwidth]{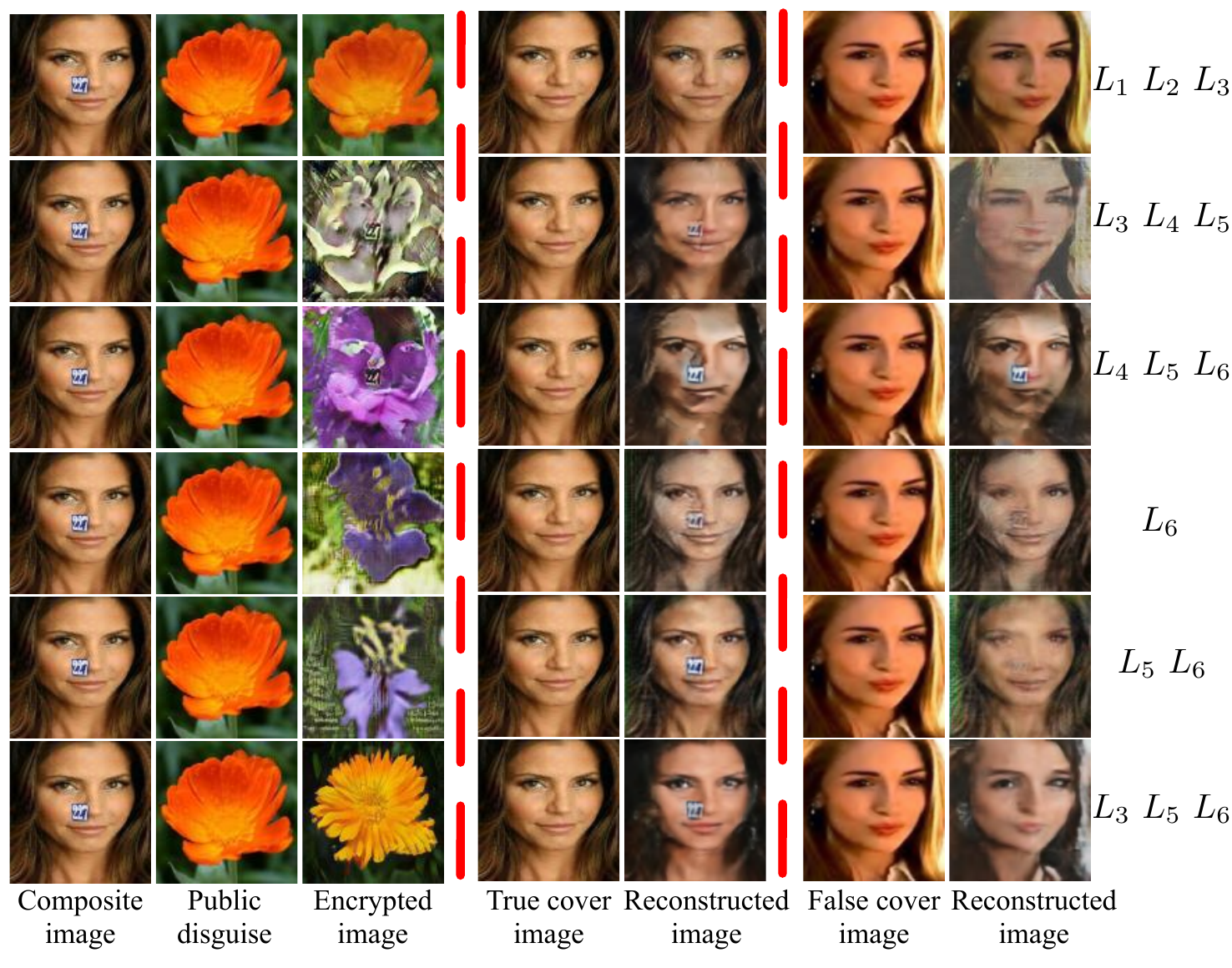}
	\caption{The ablation study of using different layer combinations for our EncryptGAN.} 
	\label{fig:ablation study} 
\end{figure}

In this section, to investigate which combination of layers could lead to better image encryption and decryption performance, we conduct experiments by using different layers for image steganography. Experimental results are shown in Figure~\ref{fig:ablation study} and Table~\ref{table:ablation}. As shown in Figure~\ref{fig:ablation study}, if we use shallow layers such as $L_1$, $L_2$ and $L_3$, the disentangled representations mainly focus on the detailed part information, and the model fails to reconstruct the required message information. If we only consider the high layer representations, e.g., $L_6$ only, we can still recover the information with a correct cover image (private key), but fail to hide the original cover images with an incorrect key, which may lead to an exposure of private key. Thus, we achieve the best performance when we choose $L_3$, $L_5$ and $L_6$, which includes both low-level and high-level information, as references to produce key pairs.

\begin{table}
	\caption{The quantitative comparison using different layer combinations for image encryption task on flower$\leftrightarrow$face translation.}
	\label{table:ablation}
	\centering
	\resizebox{\textwidth}{!}{
	\begin{tabular}{c c c c c c c c c}
		\toprule
		Methods &  FID $\downarrow$ & Encryption(LPIPS $\uparrow$ / MSE $\uparrow$) & MSE $\downarrow$ & RMSE $\downarrow$ & PSNR $\uparrow$ & SSIM $\uparrow$ & LPIPS $\downarrow$ & Security(LPIPS $\uparrow$ / MSE$\uparrow$)  \\
		\midrule
		$L_{1},L_{2},L_{3}$    & \textbf{41.1943} & 0.4105 / \textbf{0.1257} & 0.1290 & 0.3541 & 7.8553 & 0.8197 & 0.4058 & 0.4249 / 0.1365\\
		$L_{3},L_{4},L_{5}$ & 210.3436 &0.3259 / 0.1169 &0.0159 & 0.1131 & \textbf{18.5142} & 0.9694 & \textbf{0.1438} & \textbf{0.7116} / 0.1132\\
		$L_{4},L_{5},L_{6}$ & 139.7629 & 0.3981 / 0.0857 & 0.0187 & 0.1299 & 16.8173 & 0.9715 & 0.2291& 0.4182 / 0.0731\\
		$L_{6}$  & 321.4196  & 0.3034 / 0.0647 & 0.0184 & 0.1337 & 16.3136 & 0.9782 & 0.2527 & 0.3634 / 0.0386\\
		$L_{5},L_{6}$ & 204.7240 & 0.3825 / 0.0613 & \textbf{0.0125} & \textbf{0.1094} & 18.1165 & \textbf{0.9860} & 0.2103& 0.3934 / 0.0894\\
		$L_{3},L_{5},L_{6}$(EncryptGAN) & 107.9238 & \textbf{0.4141} / 0.0765  & 0.0139 & 0.1156 & 17.5992 & 0.9825 & 0.2414 & 0.5191 / \textbf{0.1379}\\
		\bottomrule
	\end{tabular}}
\end{table}

\subsection{Complex Information Encryption}
\begin{figure}[!t]
	\centering   
	\includegraphics[width=\textwidth]{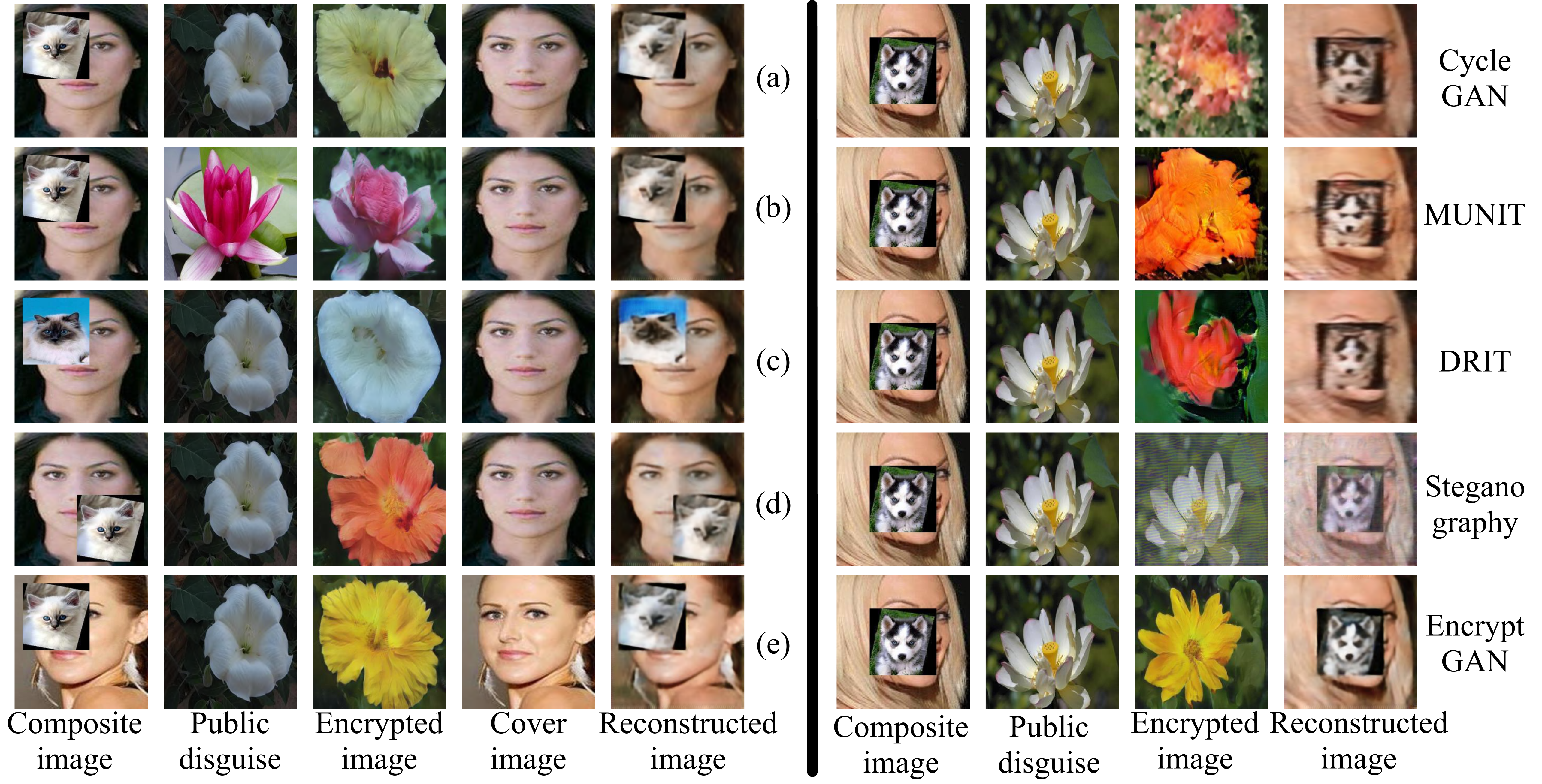}
	\caption{{\bf Left:} The effect under different circumstances similar to Figure~\ref{fig:results}. {\bf Right:} Visual comparison results of different methods.} 
	\label{fig:catdog} 
\end{figure}

For this task, we aim to encrypt more complex information such as a cat or dog image. Here we resize the cat/dog images to $128\times 128$ and paste in them to the cover image to create a $256\times 256$ composite image. As shown in Figure~\ref{fig:catdog} (Left), if we change the public disguise image, the same cover image of (a) could be encrypted to different output in (b), and if we place another message image to the same position, and the encrypted image could also have different expression such as color in (c). As for the position, if we change the position to paste in the message image, the color and shape of encrypted image will change significantly in (d). Finally, we change the cover image to carry the same message image used in (a) and (b), and successfully get the message image in the reconstructed image of (e). Please note that the synthesized encrypted image is also different in the $3_{rd}$ column of (e). From above experimental results, it is clear that our EncryptGAN works well with different disguise images, different message images as well as different positions for message images. We also exhibit the visual comparison results of different methods in Figure~\ref{fig:catdog} (Right). 

\section{Discussion and Conclusion}

We proposed a domain translation framework that achieves image asymmetric encryption by combining the public-key cryptography system, which is the first such architecture using GAN. Our EncryptGAN can achieve a semantic level message steganography and perform better than concurrent image-to-image translation as well as image steganographic methods.  We devised ablation studies for our method and performed a comprehensive and specific analysis. Comparing with deep steganography~\cite{baluja2017hiding}, which can be considered as a symmetric encryption approach for hiding information in a pixel level, our method hides messages in a semantic level under an asymmetric encryption framework. A weakness of our method is that the message we wish to deliver should be always smaller than the cover image. Thus, the pixel-wise message transmission rate of our model would be lower than the deep steganography, which can hide messages in a full scale.



\bibliographystyle{unsrtnat}

\bibliography{neurips_2019}

\end{document}